\documentclass[a4paper,11pt]{article}
\usepackage{pos}
\usepackage{setspace}

\usepackage[separate-uncertainty=true,per-mode=symbol]{siunitx}

\newcommand{\refeq}[1]{Eq.~(\ref{#1})}

\newcommand{\reffig}[1]{Fig.~\ref{#1}}

\newcommand{\refsec}[1]{Section~\ref{#1}}

\newcommand{\refref}[1]{Ref.~\cite{#1}}

\newcommand{\dEdX}{d$E$/d$X_{1500}$}
\newcommand{\sibyll}{Sibyll 2.1}
\newcommand{\epos}{EPOS-LHC}
\newcommand{\qgsjet}{QGSJet-II.04}

\title{Testing Hadronic Interaction Models with Cosmic Ray Measurements at the IceCube Neutrino Observatory}
\ShortTitle{Hadronic interaction model Studies with IceCube}

\author{The IceCube Collaboration \\{\normalsize \normalfont(a complete list of authors can be found at the end of the proceedings)}}

\emailAdd{stef.verpoest@ugent.be}

\abstract{

The IceCube Neutrino Observatory provides the opportunity to perform unique measurements of cosmic-ray air showers with its combination of a surface array and a deep detector. Electromagnetic particles and low-energy muons ($\sim$GeV) are detected by IceTop, while a bundle of high-energy muons ($\gtrsim$400 GeV) can be measured in coincidence in IceCube. Predictions of air-shower observables based on simulations show a strong dependence on the choice of the high-energy hadronic interaction model. By reconstructing different composition-dependent observables, one can provide strong tests of hadronic interaction models, as these measurements should be consistent with one another. In this work, we present an analysis of air-shower data between 2.5 and \SI{80}{\peta\eV}, comparing the composition interpretation of measurements of the surface muon density, the slope of the IceTop lateral distribution function, and the energy loss of the muon bundle, using the models Sibyll 2.1, QGSJet-II.04 and EPOS-LHC. We observe inconsistencies in all models under consideration, suggesting they do not give an adequate description of experimental data. The results furthermore imply a significant uncertainty in the determination of the cosmic-ray mass composition through indirect measurements.  \\

\vspace{4mm}
{\bfseries Corresponding authors:}
Stef Verpoest$^{1*}$, Dennis Soldin$^{2}$, Sam De Ridder$^{1}$\\
{$^{1}$ \itshape Dept. of Physics and Astronomy, University of Gent, B-9000 Gent, Belgium}\\
{$^{2}$ \itshape Bartol Research Institute and Dept. of Physics and Astronomy, University of Delaware, Newark, DE 19716, USA}\\[4mm]
$^*$ Presenter

\FullConference{37$^{\rm{th}}$ International Cosmic Ray Conference (ICRC 2021)\\
		July 12th -- 23rd, 2021\\
		Online -- Berlin, Germany}

}

\begin{document}
\maketitle

\section{Introduction}\label{sec:info}
Above \SI{100}{\TeV}, the flux of cosmic rays is too small to be studied adequately with balloon- or satellite-based experiments. Instead, large detector arrays at the Earth's surface are used to sample at a specific atmospheric depth the air showers produced by incoming cosmic rays interacting near the top of the atmosphere. To infer the energy and mass of the primary cosmic-ray nucleus from such measurements, one needs to rely on air-shower simulations. Important in these simulations is an accurate description of the high-energy hadronic interactions which govern the shower development in the atmosphere. Several hadronic interaction models exist that have been tuned to the data from accelerator experiments. Because these experiments are limited in the center-of-mass energy they can reach and the forwardness of the particles they can detect, these models rely on extrapolations, leading to uncertainties in interaction properties, such as cross-sections and multiplicities, which eventually leads to uncertainties in the simulated air-shower properties. This is most prominent in the muonic component of air showers, as muons are produced by decaying hadrons and thus carry information about the hadronic physics in the shower. A discrepancy between the number of muons in simulation and data has been established by a large number of experiments, and is known as the "Muon Puzzle"~\cite{Dembinski:2019uta}. The potential mismodeling of the hadronic interactions furthermore complicates the determination of the mass composition in the energy range above \SI{100}{\tera \eV}, where direct measurement of cosmic rays is not possible.

The IceCube Neutrino Observatory~\cite{Aartsen:2016nxy} provides a unique opportunity to study these hadronic interaction models, as its combination of a detector buried deep in the Antarctic ice sheet and a surface detector array allows it to measure multiple components of air showers at once: the electromagnetic component together with both the GeV and TeV muon components. This was first used to study hadronic interaction models by examining their consistency for different composition-sensitive observables in \refref{DeRidder:20174n}, which is continued in this work.

\begin{figure}
    \centering
    \includegraphics[width=0.45\textwidth]{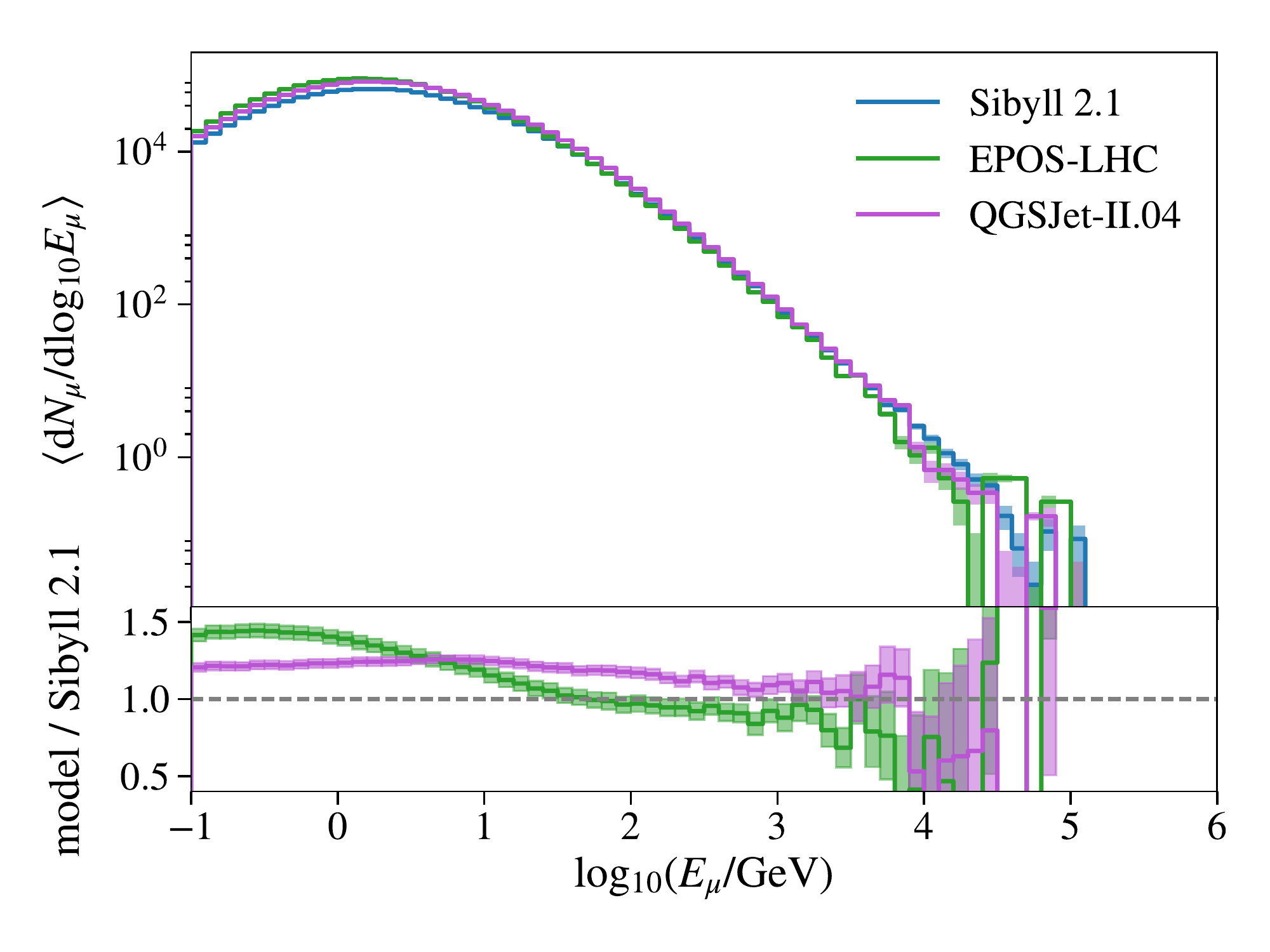}\includegraphics[width=0.45\textwidth]{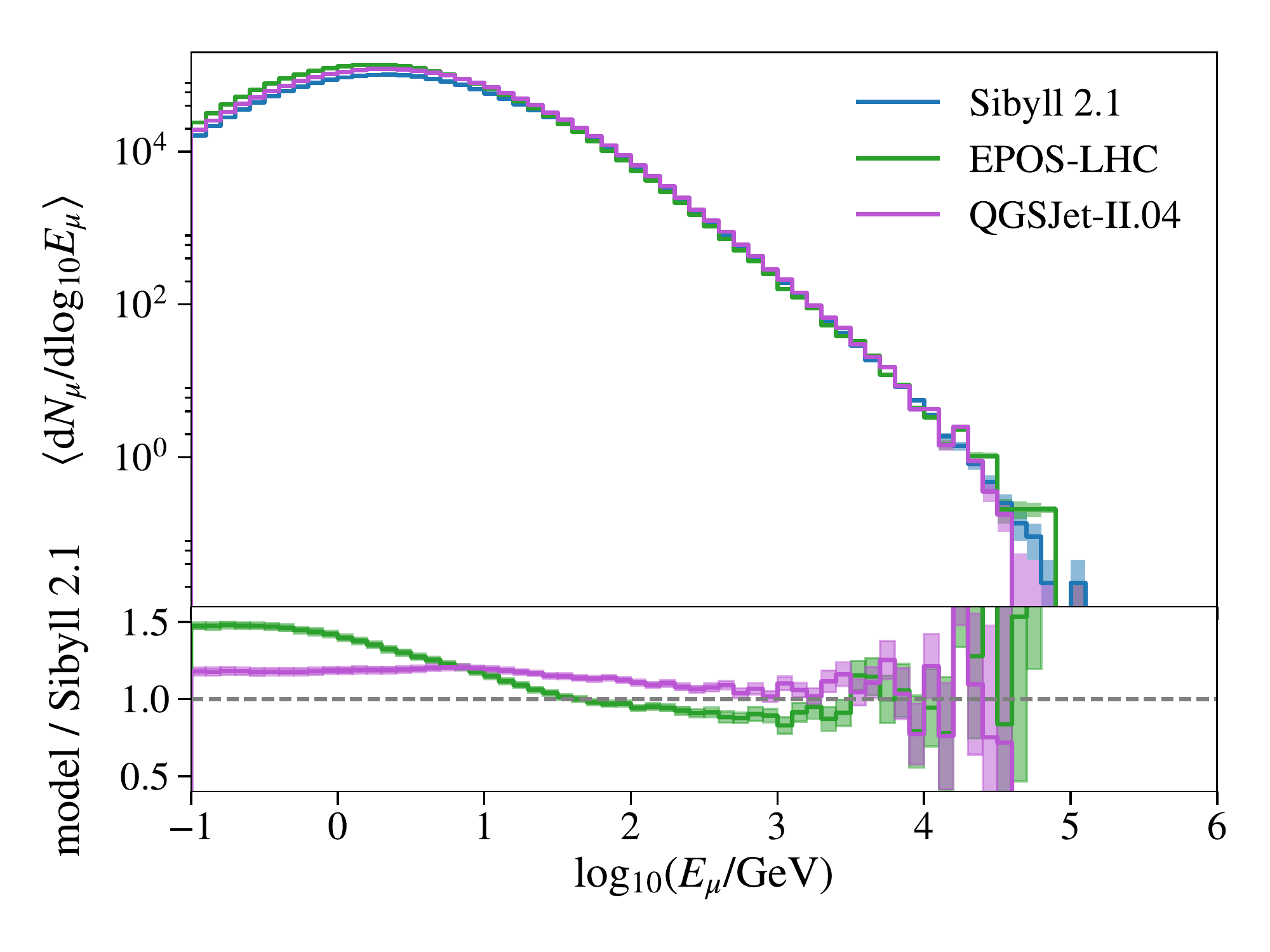}
    \vspace{-0.5em}
    \caption{Muon spectrum in vertical proton (left) and iron (right) showers with primary energy around \SI{10}{\peta \eV} for \sibyll{}, \qgsjet{} and \epos{}.}
    \label{fig:muon_spectra}
\end{figure}

\begin{figure}
    \centering
    \includegraphics[width=0.45\textwidth]{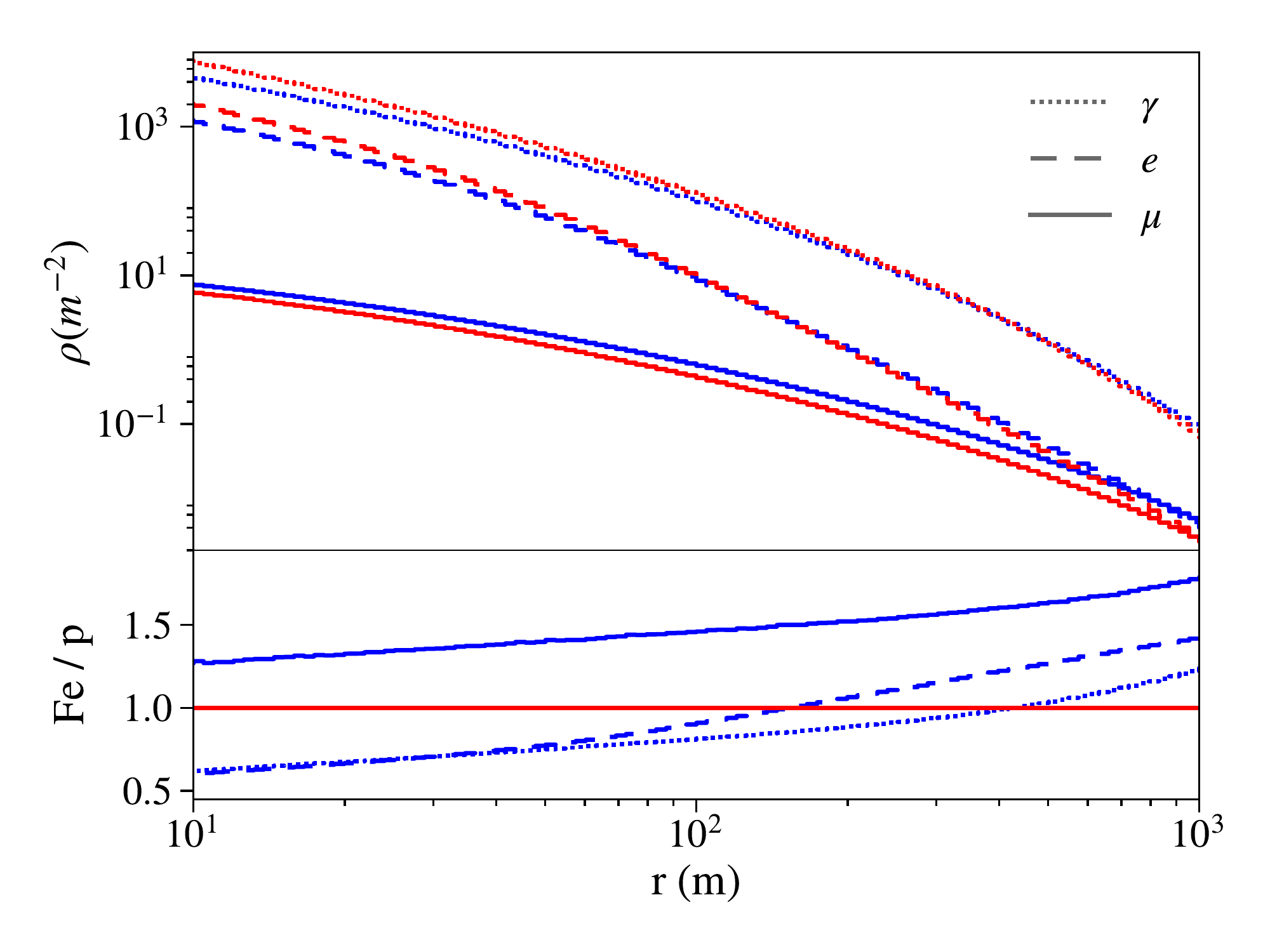}\includegraphics[width=0.45\textwidth]{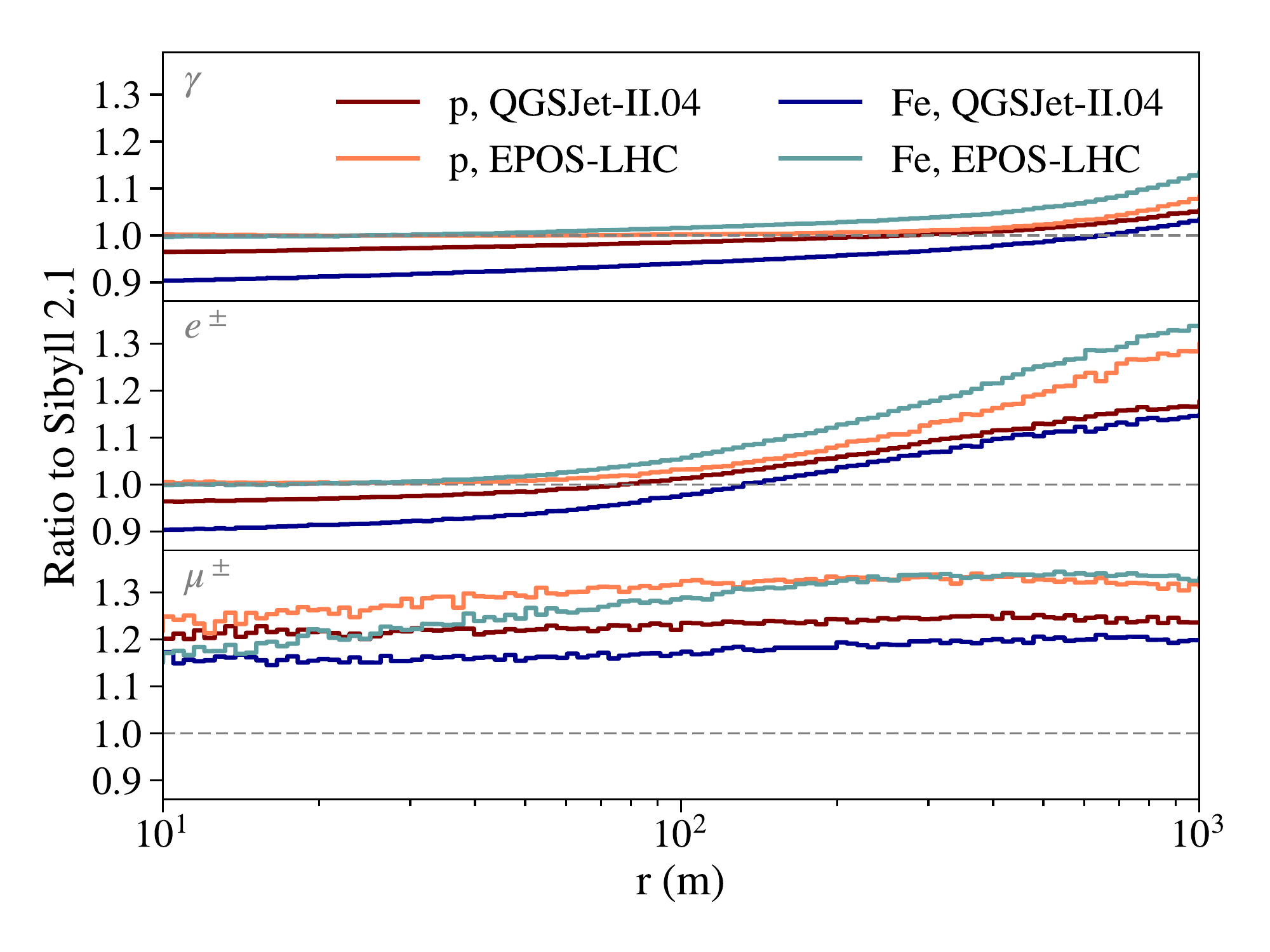}
    \vspace{-0.5em}
    \caption{Left: Lateral distribution functions of photons, electrons/positrons, and muons/antimuons in vertical proton (red) and iron (blue) showers around \SI{10}{\peta \eV} for \sibyll{}. Right: Ratio of the different LDFs in \qgsjet{} and \epos{} compared to \sibyll{}.}
    \label{fig:particle_LDF}
\end{figure}

Cosmic-ray air-shower simulations used in this work are performed with CORSIKA~\cite{CORSIKA_Heck}, using an average April South Pole atmospheric profile, and an observation level of \SI{2834}{\m} a.s.l. The high-energy hadronic interaction models considered are \sibyll{}~\cite{Ahn:2009wx}, \qgsjet{}~\cite{Ostapchenko:2010vb}, and \epos{}~\cite{Pierog:2013ria}. These last two are referred to as post-LHC models, because they are tuned to LHC data, while \sibyll{} is a pre-LHC  model\footnote{The post-LHC model Sibyll 2.3d is not included because of the absence of sufficient simulations at the time of writing.}. Low-energy interactions below \SI{80}{\GeV} are handled by FLUKA 2011.2c~\cite{BOHLEN2014211,BATTISTONI201510}. Before discussing the actual measurements, we highlight some differences between the high-energy models which are relevant for the case of IceCube. \reffig{fig:muon_spectra} shows the muon spectra in air showers as predicted by simulations using the different models under consideration. Both post-LHC models predict a larger number of GeV muons compared to \sibyll{}, while air showers simulations using \qgsjet{} contain more TeV muons than \sibyll{}, and those using \epos{} contain less. \reffig{fig:particle_LDF} shows the lateral distribution functions (LDF) of photons, (anti-)electrons, and (anti-)muons for \sibyll{}, as well as the changes in the post-LHC models. In addition to the increase of muons over the entire distance up to \SI{1}{\km} away from the core, we note that the LDFs of the electromagnetic (EM) component are less steep.

\section{Cosmic ray measurements with IceCube}\label{ref:icecube}
\begin{figure}
    \centering
    \includegraphics[height=0.48\textwidth]{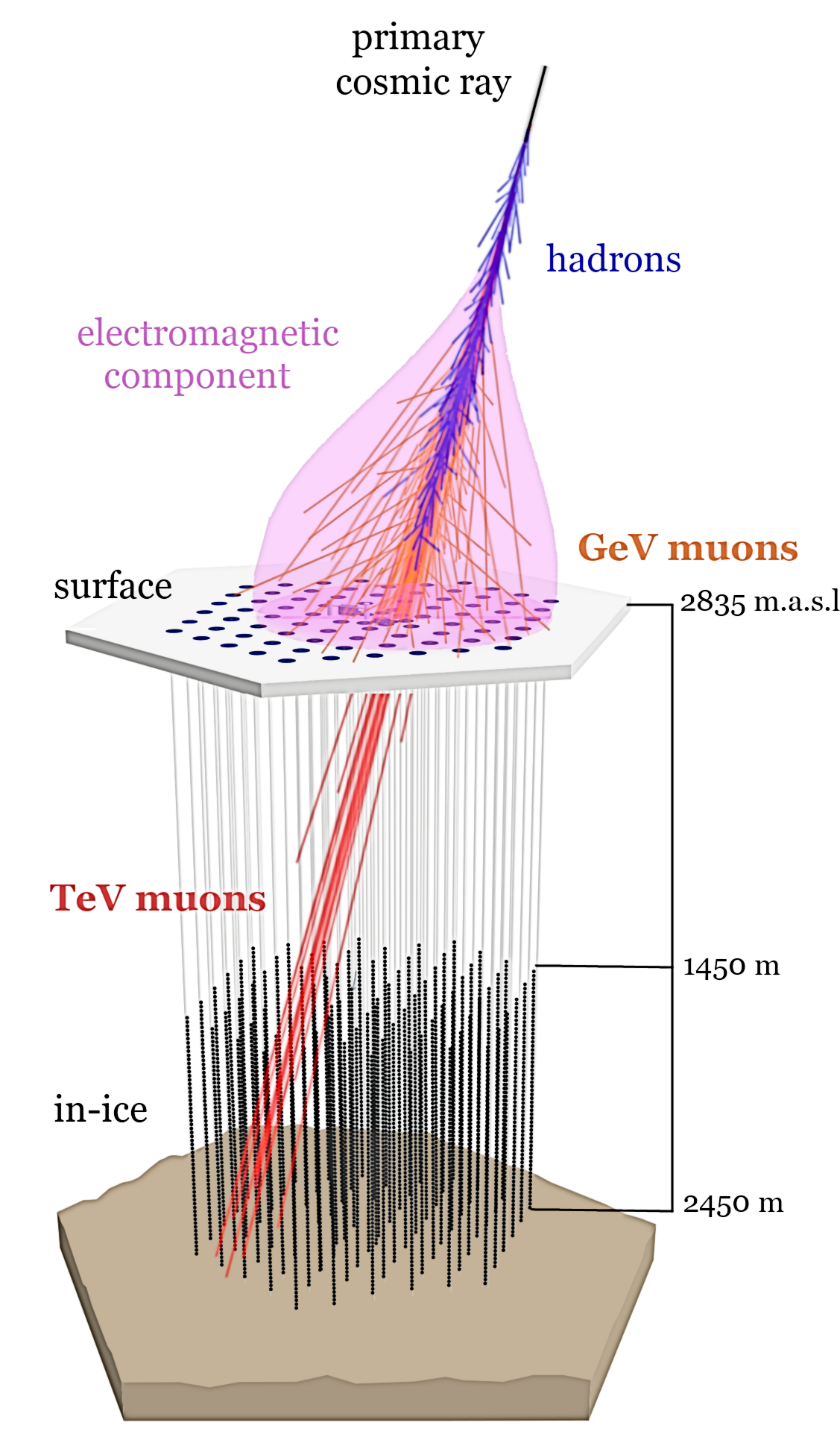}\hspace{1cm}\includegraphics[height=0.43\textwidth]{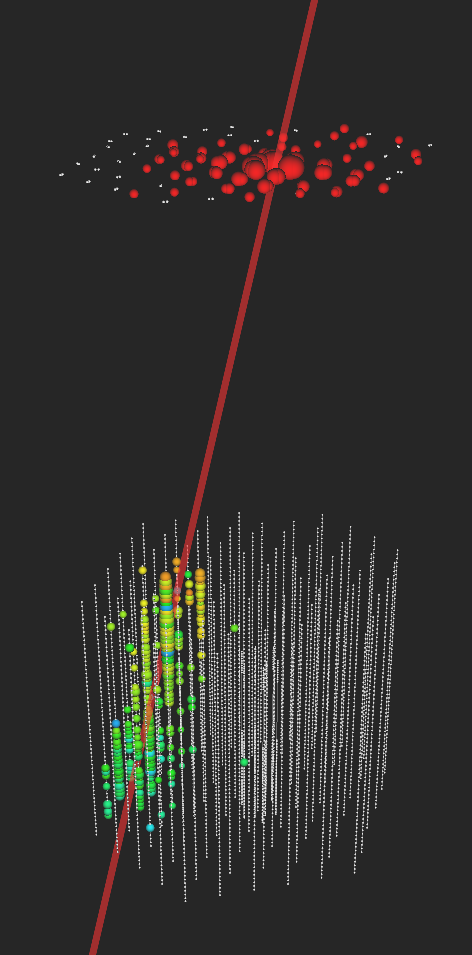}
    \caption{Left: Schematic view of an air shower observed with the IceCube Neutrino Observatory. Right: Event display of a simulated \SI{10}{\peta \eV} proton shower, coincident between IceTop and IceCube. The colour represents the time distribution of the signals, from red (early) to blue (late).}
    \label{fig:array}
\end{figure}

The IceCube Neutrino Observatory, shown in \reffig{fig:array}, consists of a surface air-shower array, IceTop~\cite{IceCube:2012nn}, and a deep in-ice detector, IceCube~\cite{Aartsen:2016nxy}, located at the geographical South Pole. IceCube instruments a volume of one cubic kilometer between depths of \SI{1450}{\m} and \SI{2450}{\m} with 5160 Digital Optical Modules (DOMs), which detect Cherenkov light generated by charged particles in the ice. The DOMs are attached to vertical strings on a triangular grid with a horizontal spacing of \SI{125}{\m} and a vertical spacing of \SI{17}{\m}. IceTop consists of 81 stations, deployed over \SI{1}{\kilo \m \squared}, following approximately the same grid as the IceCube strings. The stations consist of two ice-Cherenkov tanks separated by \SI{11}{\m}, each containing two DOMs with different gain settings to cover a large dynamic range.
The IceTop geometry is optimized for the detection of air showers from cosmic rays with primary energies between \SI{1}{\peta \eV} and \SI{1}{\exa \eV}, measuring mainly the electromagnetic and low-energy muon component of the shower. The high altitude of the array of \SI{2835}{\m} corresponds to an average atmospheric depth of \SI{692}{\g \per \cm \squared}, close to the depth of shower maximum. IceTop signals are expressed in units of `vertical equivalent muons' (VEM), the typical charge deposited by a single muon passing vertically through a tank~\cite{IceCube:2012nn}.  Because of snow accumulation on top of the IceTop tanks, the energy at which the detector becomes fully efficient increases with time. For the data season under consideration in this work, May 2012 to May 2013, the energy threshold is $\log_{10} (E_0 / \si{\GeV}) = 6.4$. Muons with energies $\gtrsim \SI{400}{\GeV}$ are able to propagate through the ice and leave a signal in IceCube. Using the two detectors in coincidence allows us to measure both the low- and high-energy muons in the shower, as well as the EM component. The muons provide a strong sensitivity to the primary mass, while the EM part is crucial to reconstruct the direction and primary energy of the shower. For this study, we focus on vertical showers ($\cos \theta > 0.95$) with quality cuts which ensure that the shower core is contained in IceTop and the high-energy muon bundle passes through IceCube. Below we describe the reconstructions applied to the data which yield the observables used in this work.

\subsection{IceTop LDF}
\begin{figure}
    \centering
    \includegraphics[width=0.33\textwidth]{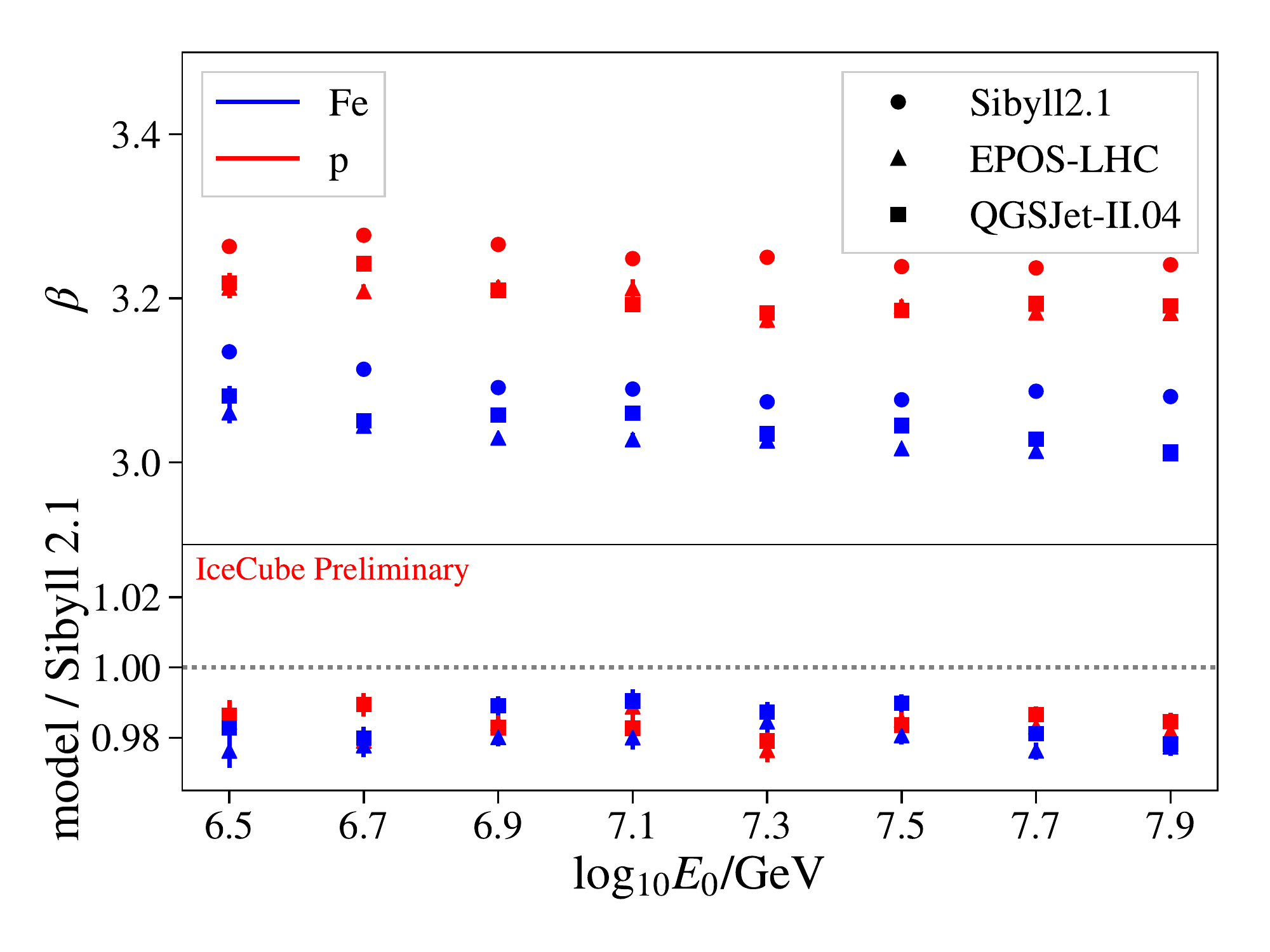}\hfill\includegraphics[width=0.33\textwidth]{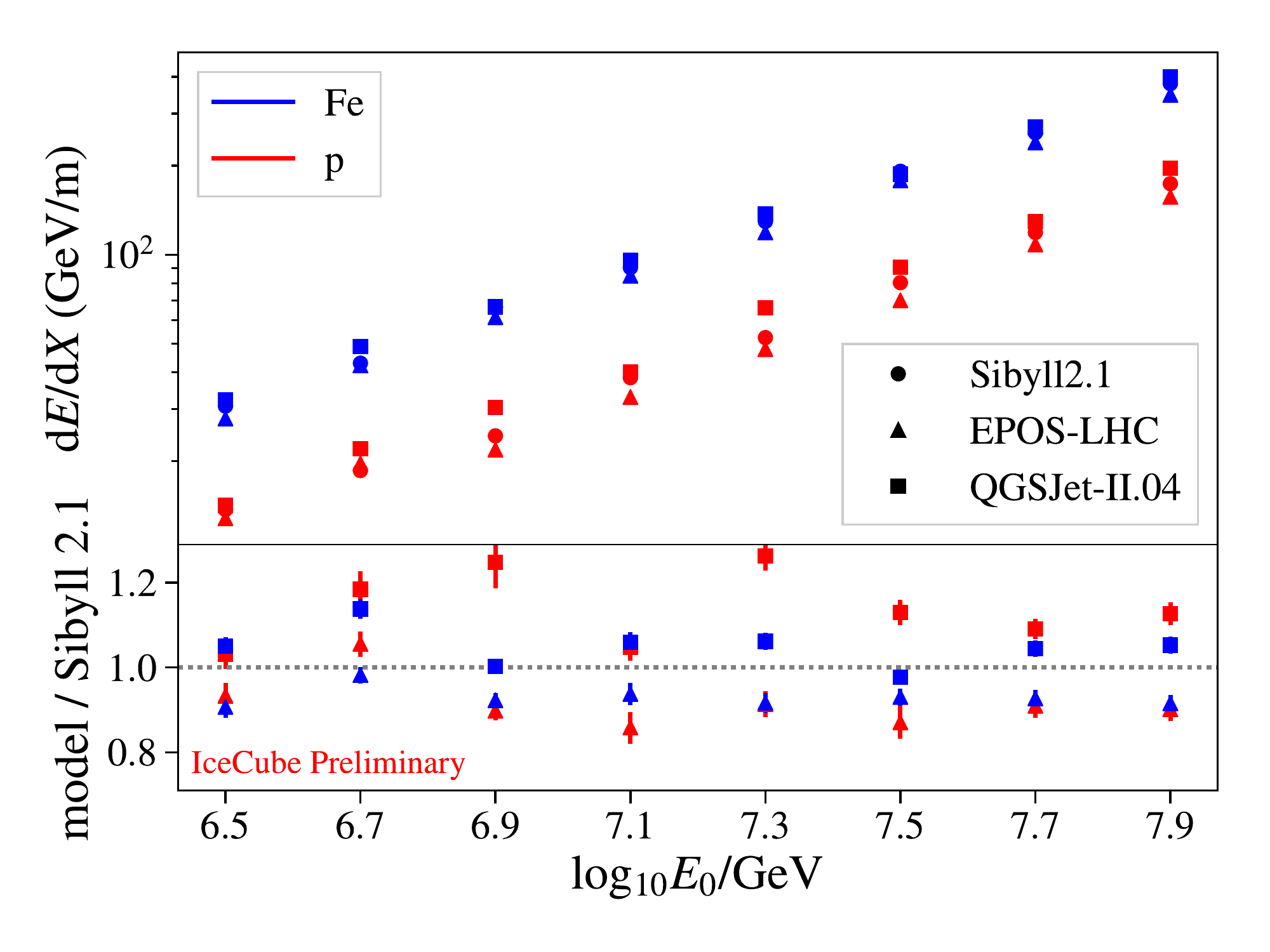}\hfill\includegraphics[width=0.33\textwidth]{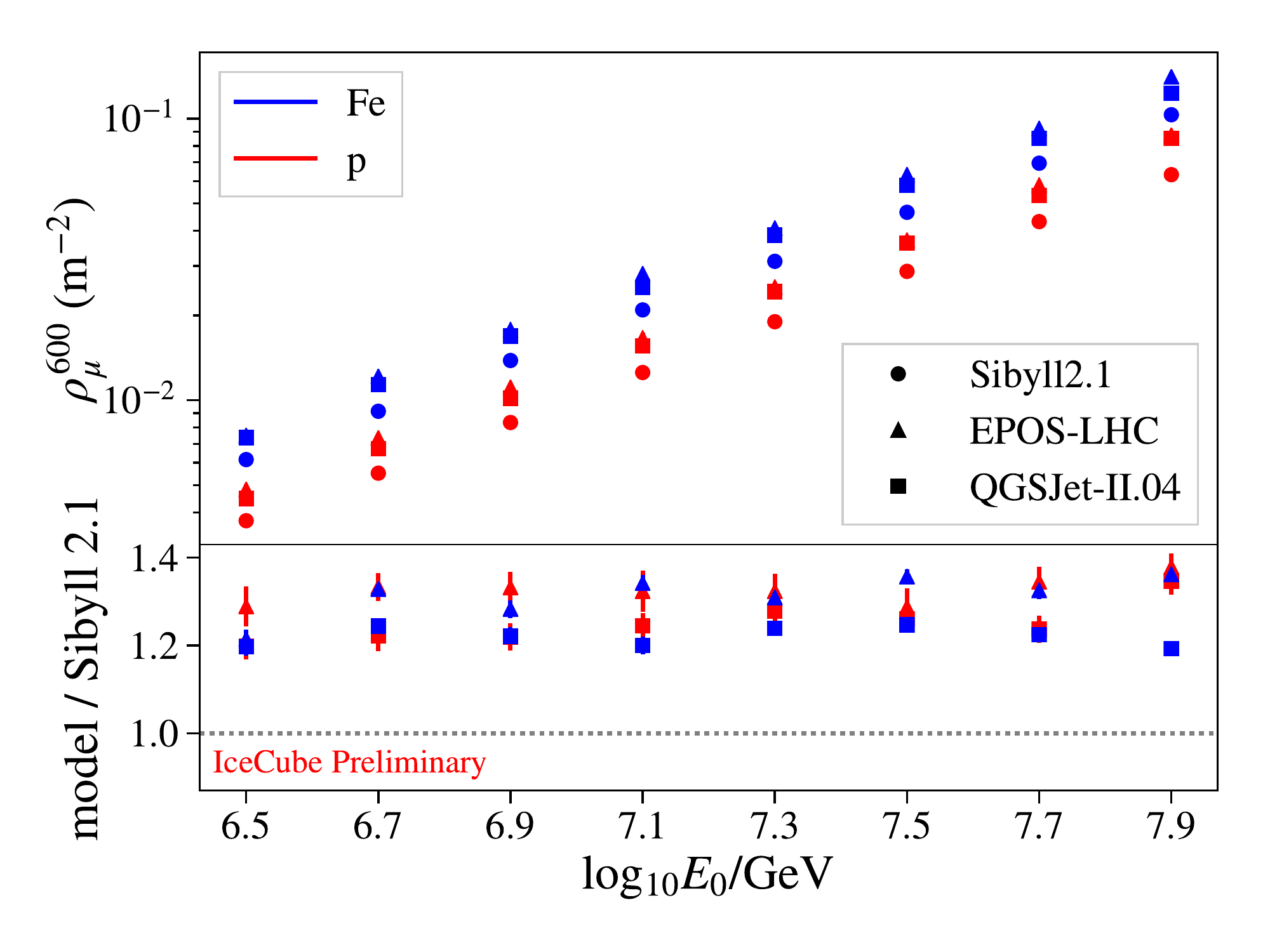}
    \caption{$\beta$, $dE/dX$, and $\rho_{\mu, 600}$, as described in the text, versus primary energy for vertical proton and iron showers. The bottom of the plot shows the ratio of the post-LHC models with respect to Sibyll 2.1.}
    \label{fig:models_var}
\end{figure}

The standard air-shower reconstruction method is performed on the IceTop signals. This is a maximum-likelihood technique which finds the best-fit shower direction and core position (as well as other parameters which are discussed later in this text) based on the charge and time of the signals which survive various noise removal algorithms~\cite{IceCube:2012nn}. As part of this reconstruction, the lateral distribution function given by
\begin{equation}
    S(R) = S_{125} \left(\frac{R}{\SI{125}{\m}}\right)^{-\beta-0.303\log_{10}\left(\frac{R}{\SI{125}{\m}}\right)} \exp\left(\frac{-d_{\mathrm{snow}}}{\lambda \cos \theta}\right),
\end{equation}
is fitted to the signals in the tanks, where $S$ is the measured charge at lateral distance $R$. The last term takes into account the attenuation of the signal due to the accumulation of snow with height $d_{\mathrm{snow}}$ on the tank. The snow mostly absorbs EM particles and not muons, so an effective attenuation length $\lambda$ is used which corrects the total tank signal. It is found to be \SI{2.25\pm0.20}{\m} for the data we consider. The two free parameters of the IceTop LDF are $\beta$, which defines its slope, and $S_{125}$, the signal at \SI{125}{\m}. Because IceTop is located close to the depth of shower maximum, the shower size parameter $S_{125}$ is determined mainly by the abundant number of EM particles in the shower and provides an accurate estimate of the primary energy of the cosmic ray with minimal model dependence~\cite{Aartsen:2013wda, IceCube:2019hmk}. The slope parameter $\beta$ is sensitive to the cosmic-ray mass, having lower values for heavier primaries, as shown in \reffig{fig:models_var}. It was verified through simulation that this dependence is a combination of two effects, relating to both the EM particles and muons measured in IceTop. First, showers from heavier primaries develop earlier in the atmosphere so that at the observation level the particles have spread out more than for lighter primaries (they have a higher shower age). Secondly, these showers also contain a larger number of muons, which further decreases the value of $\beta$, as the muon LDF is more flat than the EM LDF. The figure also shows lower values of $\beta$ in the post-LHC models, consistent with the increase of muons and flattening of the EM LDFs shown in \reffig{fig:particle_LDF}.

\subsection{Muon bundle energy loss}
The reconstructed air-shower direction and position from IceTop are used as a seed track for the reconstruction of the signal deposited by the muon bundle in IceCube. Hits that are not related to this track in time and space are removed, after which a segmented reconstruction of deposited energy along the track is performed, as described in \refref{Aartsen:2013vja}. The resulting energy loss profile is fitted to find the energy loss at a slant depth of \SI{1500}{\m}, referred to as \dEdX{}. This parameter is strongly correlated with the number of high-energy muons in a shower~\cite{IceCube:2019hmk}, with heavier primaries having larger energy loss values, as shown in \reffig{fig:models_var}. We also see higher values of \dEdX{} for \qgsjet{} compared to \sibyll{}, and lower values for \epos{}, consistent with the muon spectra of \reffig{fig:muon_spectra}.

\subsection{Density of GeV muons}
The density of GeV muons $\rho_\mu$ in IceTop is reconstructed using the method of Refs.~\cite{Gonzalez:2019epd,Soldin:2021GeVmuons}, which relies on the characteristic signal produced by a muon which passes vertically through a tank. Far from the shower axis, this can be used to statistically separate the muon signal from hits of other charged particles. This results in a statistical measurement of the mean density of GeV muons per event, derived over the entire data sample, at lateral reference distances of \SI{600}{\m} and \SI{800}{\m}. As shown in \reffig{fig:models_var}, the true muon density in simulation is higher for heavier than for lighter primaries and also higher for the post-LHC models compared to \sibyll{}.

\section{Internal consistency of models}
\begin{figure}
    \centering
    \includegraphics[width=0.45\textwidth]{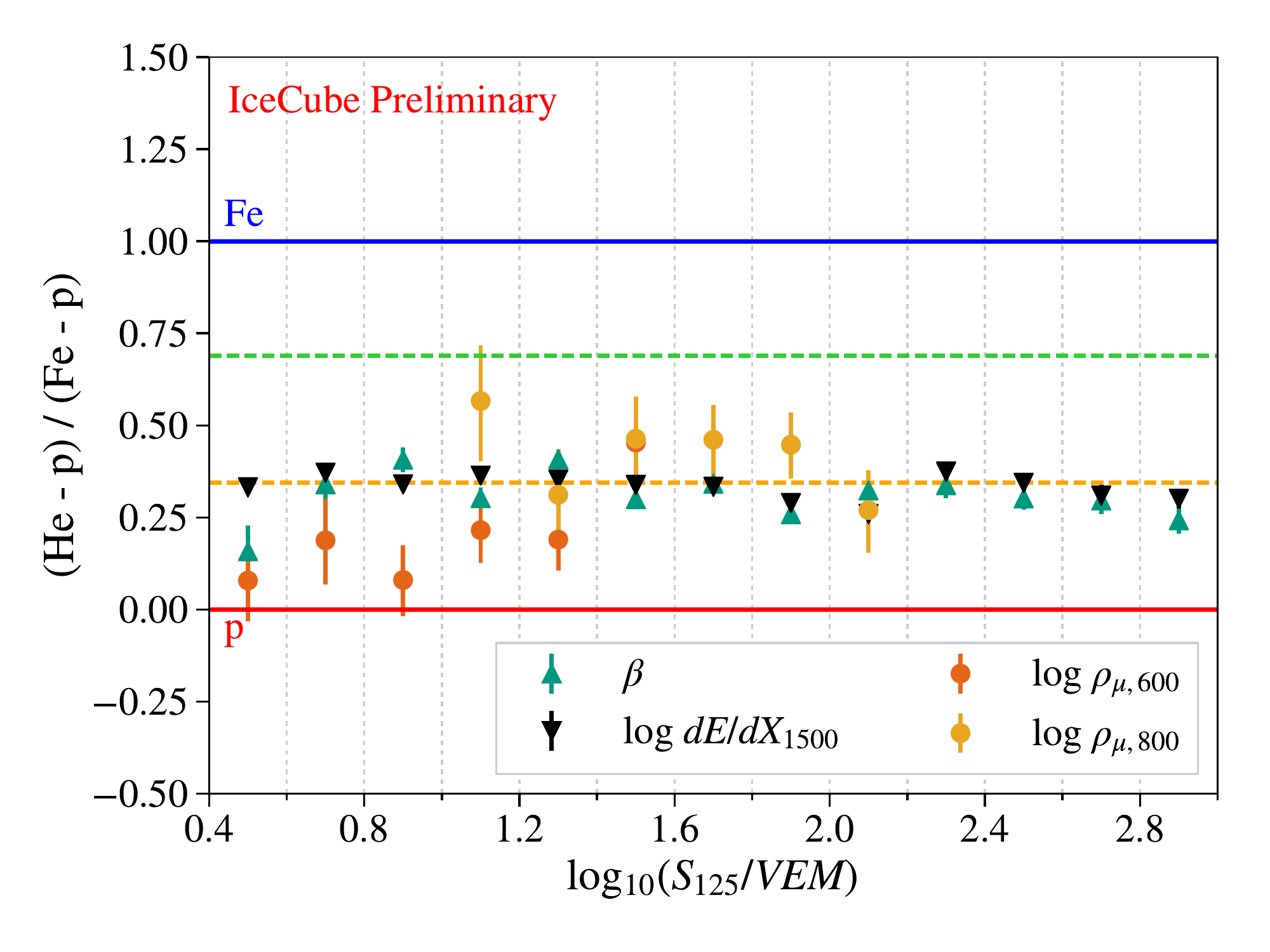}\includegraphics[width=0.45\textwidth]{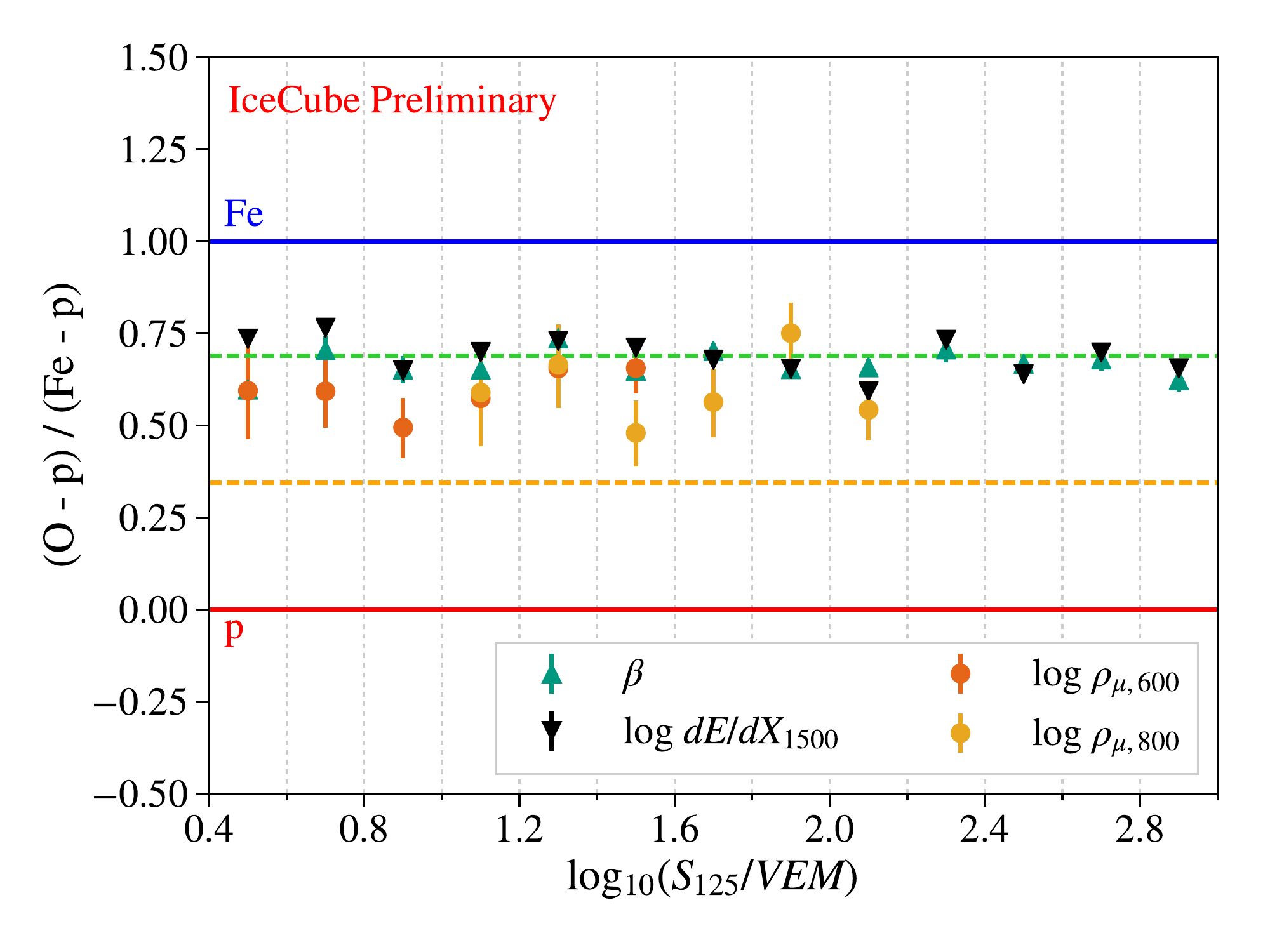}
    \vspace{-0.5em}
    \caption{Comparison of helium (left) and oxygen (right) simulation to proton and iron simulation using Sibyll 2.1. The calculated $z$-values are given for the different observables as function of $S_{125}$. The dashed horizontal lines are drawn at $\ln 4/ \ln 56$ and $\ln 16 / \ln 56$.}
    \label{fig:test}
    \vspace{-0.5em}
\end{figure}

We test the internal consistency of air-shower simulations based on different hadronic interaction models by comparing the composition-sensitive observables described in \refsec{ref:icecube} between experimental data and simulation. The distributions are given as a function of $\log_{10} (S_{125}/\mathrm{VEM})$, using a bin width of 0.2, with a lower limit of 0.4 and an upper limit of 3 for \sibyll{} and an upper limit of 2 for \qgsjet{} and \epos{}, limited by the simulation available at the time of writing. The lower limit roughly corresponds to a primary energy of $\log_{10} (E_0 / \si{\GeV}) = 6.4$, the threshold for full efficiency, while the upper values correspond to a $\log_{10} (E_0 / \si{\GeV})$ of about 8.8 and 7.9 respectively, though the precise conversion is composition dependent. Per bin, we calculate the so-called `$z$-value'\cite{DeRidder:20174n,Dembinski:2019uta}
\begin{equation}
    z = \frac{x_{\mathrm{data}} - x_\mathrm{p}}{x_\mathrm{Fe} - x_\mathrm{p}},
    \label{eq:z}
\end{equation}
for each variable $x \in \{\beta, \ln (\mathrm{d}E/\mathrm{d}X_{1500}), \ln (\rho_{\mu,600}), \ln (\rho_{\mu,800})\}$, where $x_\mathrm{p}$ and $x_\mathrm{Fe}$ are derived from proton and iron simulations based on a given hadronic model. The $z$-value would be 0 and 1 for a pure proton and iron composition respectively and somewhere in between for a mixed composition, while values outside this range indicate a discrepancy between data and simulation. The muon densities $\rho_\mu$ are determined over the entire sample, while for the event-by-event variables $\beta$ and \dEdX{} we take the mean value in each bin. The reason to use these variables and take the logarithm for some of them, is that for variables $x \propto \ln A$, \refeq{eq:z} reduces to
\begin{equation}
    z = \frac{\ln A_\mathrm{data}}{\ln 56}.
\end{equation}
From the Heitler-Matthews model~\cite{Matthews:2005sd}, this relation is expected for the number of muons in a shower. By calculating the reconstructed $z$-values from helium and oxygen simulations, we see that this relation approximately holds for all variables under consideration here, as illustrated in \reffig{fig:test}.

\begin{figure}
    \centering
    \vspace{-1.5em}
    
    \includegraphics[width=.5\textwidth]{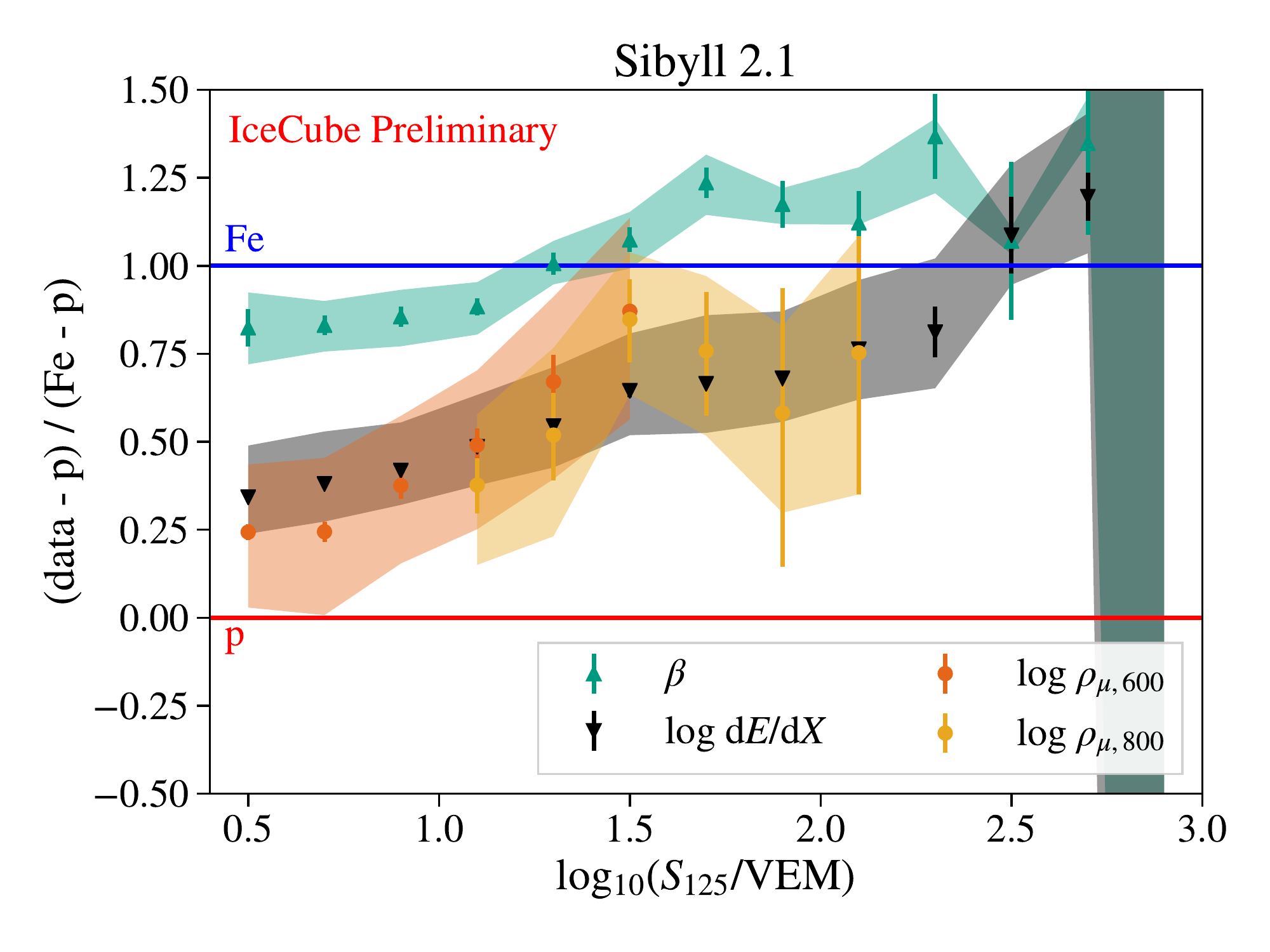}
    
    \vspace{-.3cm}\includegraphics[width=.5\textwidth]{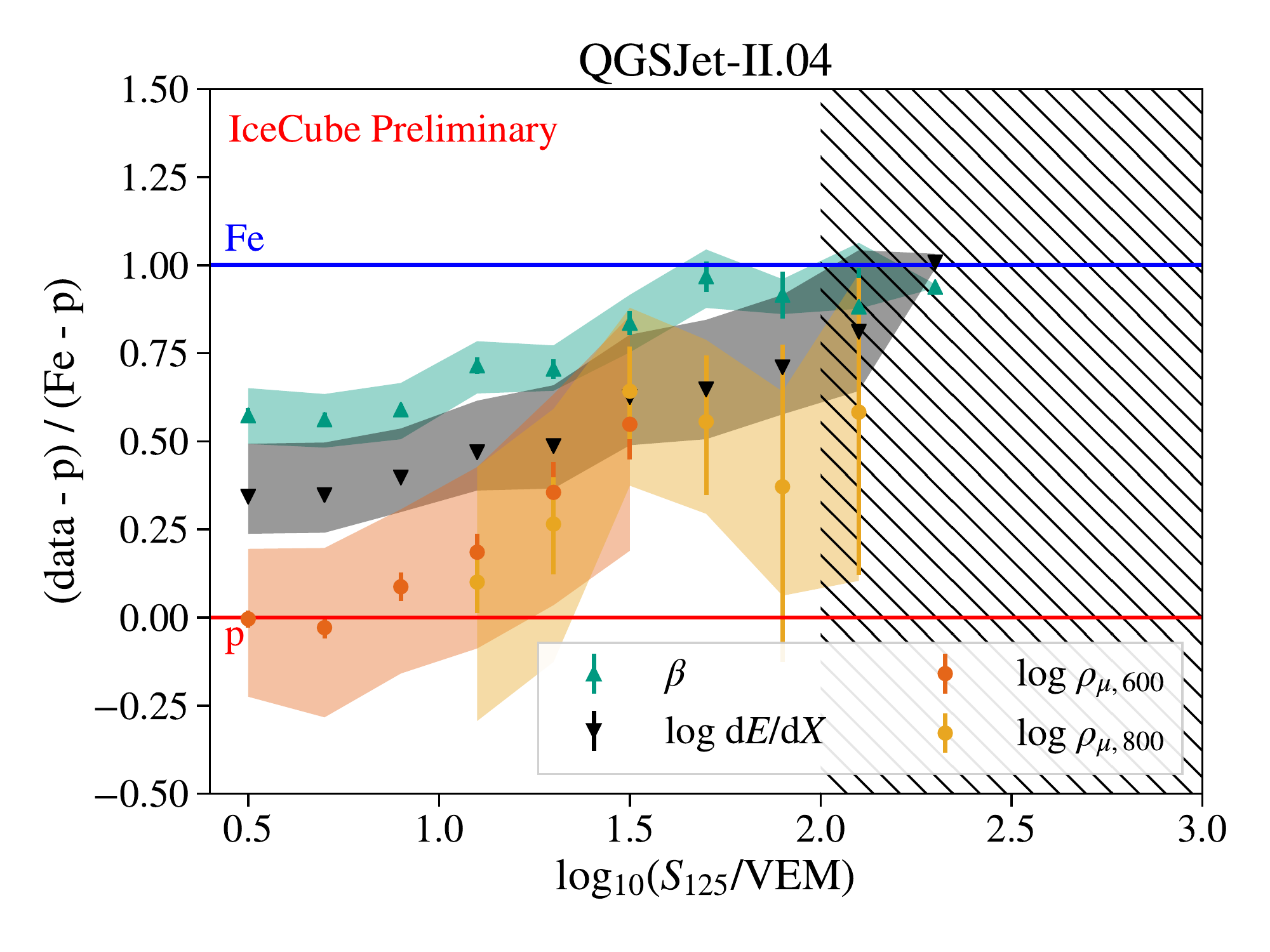}\includegraphics[width=.5\textwidth]{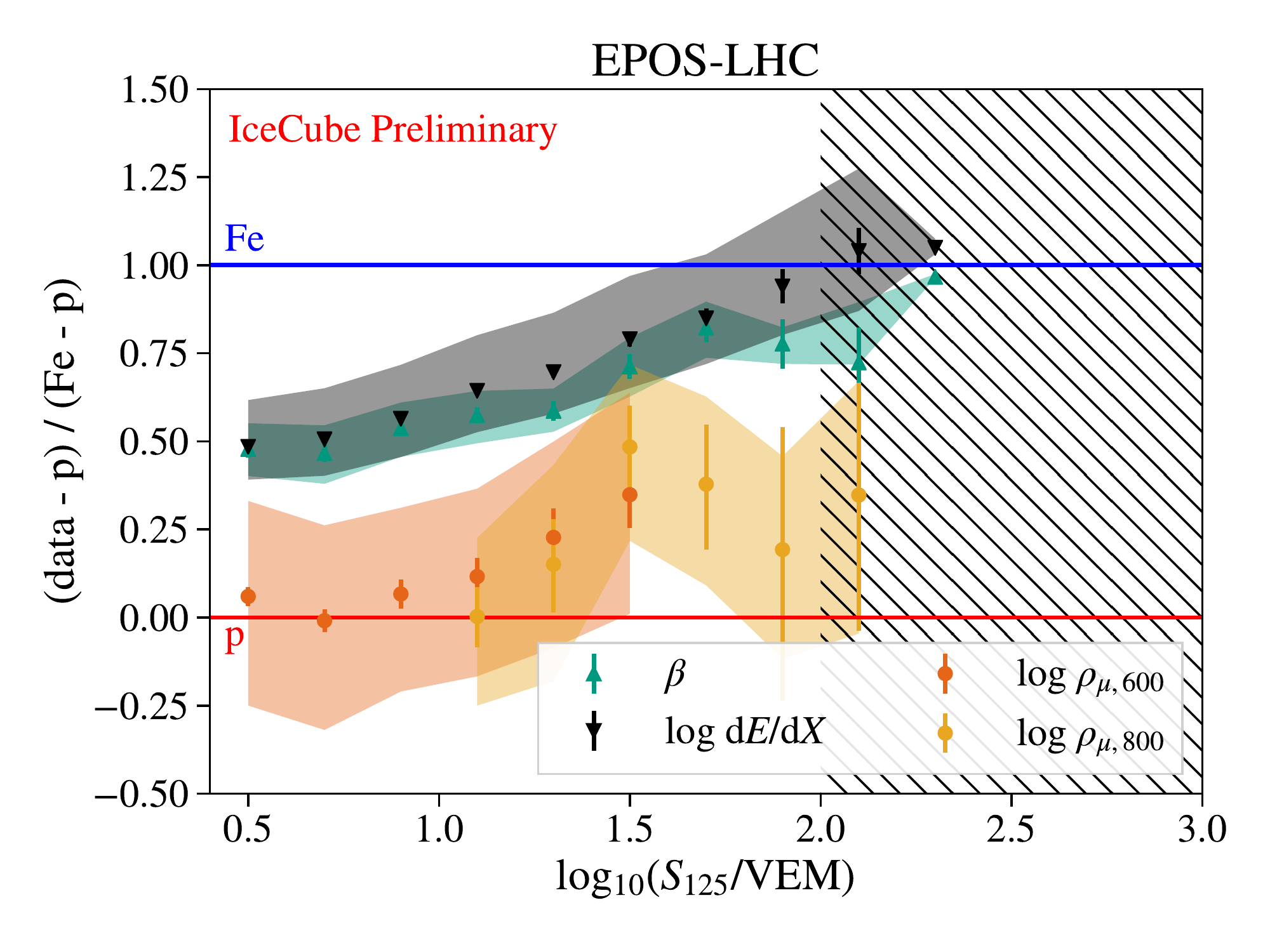}
    \vspace{-2em}
    
    \caption{Distribution of the different composition-sensitive observables as function of the primary energy estimator $S_{125}$ in proton-iron space as represented by the $z$-values. The error bars show the statistical uncertainty, while the bands represent the systematic uncertainties. Due to a limited availability of simulations, the results for \qgsjet{} and \epos{} are limited to $\log_{10} S_{125} / \mathrm{VEM} = 2$.}
    \label{fig:results}
    \vspace{-0.5em}
\end{figure}

If the simulations describe the distributions of these different variables in data correctly, the corresponding $z$-values are expected to overlap because they all follow the the same underlying mass composition. The resulting $z$-values, determined using 10\% of the experimental data between May 2012 to May 2013 and simulations based on different hadronic models, are shown in \reffig{fig:results}. The systematic uncertainties on $\beta$ result from a \SI{0.2}{\m} uncertainty on the snow attenuation length $\lambda$ and a $\pm 3\%$ uncertainty on the calibration of the charges. For \dEdX{}, this is accompanied by uncertainties on the scattering and absorption in the bulk ice, on the scattering in the refrozen ice surrounding the strings and on the efficiency of the DOMs, which combines into a total light yield uncertainty of $-12.5\%$ and $+9.6\%$~\cite{IceCube:2019hmk}. The systematic uncertainties for $\rho_\mu$ are dominated by the fits to signal distributions in the analysis method, as discussed in \refref{Gonzalez:2019epd}.

We observe that for all models the curves corresponding to the different variables increase with $S_{125}$, consistent with a composition that becomes increasingly heavy. Although the general trend is similar, there are striking inconsistencies between the variables. For \sibyll{}, we see that the LDF slope $\beta$ indicates a composition that is much heavier than for the low- and high-energy muon measurements, even going beyond iron, which may indicate that the predicted slope of the EM LDF is too steep. The muon measurements on the other hand overlap with eachother and give a consistent composition interpretation. For the post-LHC models, the muon density looks more proton-like in data because of the increase in muons in the simulations. For \qgsjet{}, we notice little overlap between all considered variables, which cover a large fraction of the area between proton and iron. For \epos{}, the values of \dEdX{} shifts up compared to \sibyll{} and \qgsjet{}, consistent with its lower number of high-energy muons. While there is acceptable agreement with $\beta$, there is a large inconsistency between the low- and high-energy muons.

\section{Conclusion and Outlook}
The IceCube Neutrino Observatory is able to perform unique studies of hadronic interactions in high-energy cosmic-ray physics with its capability to measure simultaneously the electromagnetic, GeV muon and TeV muon components of air showers. In this work, we presented tests of \sibyll{}, \qgsjet{}, and \epos{}, comparing data to proton and iron simulations for three different composition-sensitive observables. If the models give a realistic description of experimental data, the composition interpretation of all observables should be consistent. However, we observe inconsistencies between different components in the models considered, notably between the LDF slope and the low-energy muons in all models, and the low- and high-energy muons in the post-LHC models. These inconsistencies furthermore make it challenging to unambiguously determine the mass composition of cosmic rays through indirect measurements.

Inclusion of more data, increased availability of simulations, and a better understanding of systematics will advance the precision of the tests we presented. Future work improving measurements of low- and high-energy muons will provide additional input relevant for the understanding of air shower physics. The extension of the Observatory with a larger in-ice detector and new surface detectors~\cite{Aartsen:2020fgd,Haungs:2019ylq,Paul:2021IceAct} will further allow to expand this work with more complementary measurements.



\bibliographystyle{ICRC}
\bibliography{references}

\clearpage
\section*{Full Author List: IceCube Collaboration}




\scriptsize
\noindent
R. Abbasi$^{17}$,
M. Ackermann$^{59}$,
J. Adams$^{18}$,
J. A. Aguilar$^{12}$,
M. Ahlers$^{22}$,
M. Ahrens$^{50}$,
C. Alispach$^{28}$,
A. A. Alves Jr.$^{31}$,
N. M. Amin$^{42}$,
R. An$^{14}$
% ...
\\

\noindent
% ...
$^{12}$ Universit{\'e} Libre de Bruxelles, Science Faculty CP230, B-1050 Brussels, Belgium \\
% ...
$^{14}$ Department of Physics and Laboratory for Particle Physics and Cosmology, Harvard University, Cambridge, MA 02138, USA \\
% ...
$^{17}$ Department of Physics, Loyola University Chicago, Chicago, IL 60660, USA \\
$^{18}$ Dept. of Physics and Astronomy, University of Canterbury, Private Bag 4800, Christchurch, New Zealand \\
% ...
$^{22}$ Niels Bohr Institute, University of Copenhagen, DK-2100 Copenhagen, Denmark \\
% ...
$^{28}$ D{\'e}partement de physique nucl{\'e}aire et corpusculaire, Universit{\'e} de Gen{\`e}ve, CH-1211 Gen{\`e}ve, Switzerland \\
% ...
$^{31}$ Karlsruhe Institute of Technology, Institute for Astroparticle Physics, D-76021 Karlsruhe, Germany  \\
% ...
$^{42}$ Bartol Research Institute and Dept. of Physics and Astronomy, University of Delaware, Newark, DE 19716, USA \\
% ...
$^{50}$ Oskar Klein Centre and Dept. of Physics, Stockholm University, SE-10691 Stockholm, Sweden \\
% ...
$^{59}$ DESY, D-15738 Zeuthen, Germany \\
% ...

\subsection*{Acknowledgements}

\noindent
USA {\textendash} U.S. National Science Foundation-Office of Polar Programs,
U.S. National Science Foundation-Physics Division,
U.S. National Science Foundation-EPSCoR,
Wisconsin Alumni Research Foundation,
Center for High Throughput Computing (CHTC) at the University of Wisconsin{\textendash}Madison,
Open Science Grid (OSG),
Extreme Science and Engineering Discovery Environment (XSEDE),
Frontera computing project at the Texas Advanced Computing Center,
U.S. Department of Energy-National Energy Research Scientific Computing Center,
Particle astrophysics research computing center at the University of Maryland,
Institute for Cyber-Enabled Research at Michigan State University,
and Astroparticle physics computational facility at Marquette University;
Belgium {\textendash} Funds for Scientific Research (FRS-FNRS and FWO),
FWO Odysseus and Big Science programmes,
and Belgian Federal Science Policy Office (Belspo);
Germany {\textendash} Bundesministerium f{\"u}r Bildung und Forschung (BMBF),
Deutsche Forschungsgemeinschaft (DFG),
Helmholtz Alliance for Astroparticle Physics (HAP),
Initiative and Networking Fund of the Helmholtz Association,
Deutsches Elektronen Synchrotron (DESY),
and High Performance Computing cluster of the RWTH Aachen;
Sweden {\textendash} Swedish Research Council,
Swedish Polar Research Secretariat,
Swedish National Infrastructure for Computing (SNIC),
and Knut and Alice Wallenberg Foundation;
Australia {\textendash} Australian Research Council;
Canada {\textendash} Natural Sciences and Engineering Research Council of Canada,
Calcul Qu{\'e}bec, Compute Ontario, Canada Foundation for Innovation, WestGrid, and Compute Canada;
Denmark {\textendash} Villum Fonden and Carlsberg Foundation;
New Zealand {\textendash} Marsden Fund;
Japan {\textendash} Japan Society for Promotion of Science (JSPS)
and Institute for Global Prominent Research (IGPR) of Chiba University;
Korea {\textendash} National Research Foundation of Korea (NRF);
Switzerland {\textendash} Swiss National Science Foundation (SNSF);
United Kingdom {\textendash} Department of Physics, University of Oxford.

\end{document}